# Tuning the plasmon resonance of metallic tin nanocrystals in Si-based materials.

Mads Møgelmose Kjeldsen[1], John Lundsgaard Hansen[1], Thomas Garm Pedersen[2], Peter Gaiduk[3] and Arne Nylandsted Larsen[1].

[1] Department of Physics and Astronomy and iNANO, Aarhus University, Ny Munkegade 120, DK-8000 Aarhus C, Denmark, fax: +45-86120740, e-mail: madsmk@phys.au.dk.

[2] Department of Physics and Nanotechnology, Aalborg University, DK-9220 Aalborg Oe, Denmark.

[3] Physical Electronics Department, Belarusian State University, 220030 Minsk, Belarus.

**Abstract.** The optical properties of metallic tin nanoparticles embedded in silicon-based host materials were studied. Thin films containing the nanoparticles were produced using RF magnetron sputtering followed by *ex situ* heat treatment. Transmission electron microscopy was used to determine the nanoparticle shape and size distribution; spherical, metallic tin nanoparticles were always found. The presence of a localized surface plasmon resonance in the nanoparticles was observed when $SiO_2$ and amorphous silicon were the host materials. Optical spectroscopy revealed that the localized surface plasmon resonance is at approximately 5.5 eV for tin nanoparticles in $SiO_2$, and at approximately 2.5 eV in amorphous silicon. The size of the tin nanoparticles in $SiO_2$ can be varied by changing the tin content of the films; this was used to tune the localized surface plasmon resonance.

# 1 Introduction

Metallic nanoparticles are known to exhibit strong resonant absorption due to the excitation of localized surface plasmon (LSP) resonances [1-11]. The resonance energy depends on the material, size, shape and the surrounding medium of the nanoparticles; these parameters were all previously used to tune the LSP resonance [4-6]. The noble metals gold and silver have been the far most investigated nanoparticle materials due to their inertness and resonances in the visible light region. Furthermore, enhancement of photovoltaic devices and light emitting structures was also achieved using LSPs [7-9]; the principle at issue is scattering of light increases the optical path length in the device and, thereby, chance of generation of electron-hole pairs. Silicon is the predominantly used material for photovoltaic devices such as solar cells. However, a disadvantage of using metal in combination with silicon is the likely introduction of metallic defects which might give rise to mid-gap energy levels [12], and, consequently, damaging the device performance. Tin is an exception to this rule because tin is electrically neutral as a substitutional impurity in silicon, and is thus compatible with the silicon process technology [13].

Metallic tin nanoparticles have so far not been the target of optical investigations. Metallic tin, also called white tin or β-Sn, has an anisotropic, body-centered, tetragonal structure, and is the allotropic phase assumed by bulk tin at room temperature and atmospheric pressure. A minor drawback of the use of β-tin is its oxidization when in contact with air. However, the oxidation can be avoided by embedding the nanoparticles in a host material. In the present investigation, $SiO_2$ and amorphous silicon (a-Si) were chosen as host materials. Both are silicon based, and are already used in photovoltaic devices. Note that the bond enthalpy of the Si-O bond (800 kJ/mol) is significantly larger than that of the Sn-O bond (532 kJ/mol) [14]. Therefore, a $SiO_2$ host is not likely to cause oxidation of tin nanoparticles by oxygen interdiffusion. It should also be noted that $SiO_2$ is a low refractive index material ($n$=1.5), whereas a-Si is a high refractive index material ($n$=3.4 [15]) because the LSP resonance energy is correlated to the refractive index of the host material.

Structural and optical properties of metallic tin nanoparticles in $SiO_2$ and a-Si thin films were studied experimentally in the present investigation. The impact of the particle size on LSP resonance was studied for the particles in the $SiO_2$ film where the particle size can be controlled by changing the tin content.

## 2 Experimental details

Trilayered $SiO_2$/Sn/$SiO_2$ thin films were deposited on quartz substrates using RF magnetron-sputtering at a pressure of 0.002 mbar. The quartz substrates were not intentionally heated during deposition but reach nevertheless a constant temperature of ~50°C after several minutes of deposition. The elemental content of the samples was measured using Rutherford Backscattering Spectrometry (RBS) with 1.5 MeV $He^+$ particles. Four samples with linearly increasing Sn content were produced and measured by RBS. The total amounts of Sn in the samples were equivalent to a 4.5, 9.5, 13.5 and 16.0 nm thick layer of β-Sn, respectively. The thickness of every $SiO_2$ layer was aimed at 25 nm by using the sputtering rate measured on previously produced thin films; this thickness was confirmed by the RBS spectra. Subsequently, the samples were all heat treated at 450°C for 30 minutes in a tube furnace with a continuous flow of nitrogen gas. The samples were introduced into a preheated furnace, and were removed after 30 minutes and allowed to cool in the load lock while the nitrogen ambience was maintained. The samples are henceforth named sample 1 to 4 with sample 4 having the thickest Sn layer.

Using the procedure described above also trilayered thin films of Si/Sn/Si were produced on several quartz substrates and, in addition, reference samples with only an a-Si layer. The thicknesses of the individual layers of the Si/Sn/Si sample were determined to be 40 nm, 10 nm and 40 nm using RBS.

Separate pieces of samples were heat treated for 30 minutes in a nitrogen ambience at 400 and 700°C, respectively. Also a set of samples was not heat treated. Notice that the silicon thin films are expected to be amorphous after sputtering deposition and remain so when heat treated at 400°C whereas higher temperatures result in a poly-crystalline silicon film [16].

Samples for TEM investigation were prepared using standard thinning and Ar-ion milling techniques. They were characterized using a Philips CM20 TEM with a $LaB_6$ filament operated at 200 kV; images were acquired using a Gatan CCD-camera. Optical characterization was performed using a Shimatzu UV-3600 optical spectrometer. Extinction was measured with the spectrometer in a dual beam setup at normal angle of incidence. Extinction is defined as $Ext = -\log_{10}(I_T/I_0)$, where $I_0$ and $I_T$ are the intensity of incident and transmitted light, respectively. Transmission and reflection spectra were obtained by adding an integrating sphere module to the spectrometer and measuring at an 8° angle of incidence. When measuring the spectra with the integrating sphere, the transmission T represent the sum of directly transmitted light and forward scattered light, and the reflection R represent the sum of backwards scattered light and specularly reflected light. The absorption A was calculated from the equation

$$A = 1 - R - T \quad (1)$$

which is an expression of energy conservation.

## 3 Results

*3.1 Results from transmission electron microscopy*

A TEM investigation of as-deposited samples showed that the formation of nanoparticles occurs already during the deposition as demonstrated in figure 1. The minor bright areas present midway between nanoparticles in the plane-view micrographs were investigated by over- and under focusing the electron microscope; the brightness was attributed to the thickness of the film being locally decreased [17]. The formation of the nanoparticles is caused by diffusion and agglomeration of tin during the deposition.

In figure 2, a bright field TEM micrograph of the nanoparticles in sample 2 is shown in cross-sectional view. The tin nanoparticles have a circular shape, however, the dark areas between the nanoparticles on the figure might indicate that some tin is located there. The $SiO_2$ layers are visible below and above the nanoparticles, and topmost the glue of the cross-section samples preparation is distinguishable due to the brighter contrast. The image shows that these nanoparticles are surrounded by a 10 to 20 nm thick $SiO_2$ shell.

In figure 3, bright field TEM micrographs of samples one to four are displayed in plane-view (note that the scale is identical on all four images). The images reveal that the shape of the nanoparticles is circular observed in plane-view, although part of the nanoparticles in sample four were elongated to some extent. Combining the information from plane-view and cross-section TEM, we conclude that the nanocrystals of sample one to three are spherical, and that those of sample four have a slightly ellipsoidal shape without facets. A possible explanation of this ellipsoidal shape is that large particles form due to *coalescence between* smaller particles. The images in figure 3 were analyzed using ImageTool software [18] by manually specifying the diameter via two points for all nanoparticles. The nanoparticle size distributions extracted in this way are displayed to the right of the micrographs in figure 3; more than 100 nanoparticles were included for each size distribution. The size distributions demonstrate an increasing average diameter of the nanoparticles with increasing Sn content. In table 1, the average diameters with standard deviations are collected along with data derived from RBS.

Interestingly, the particle diameters in sample one and two almost follow a normal distribution (figure 3b and 3d), whereas in sample three and four the distribution of the diameters is bimodal (figure 3f and 3h). In the latter two distributions, a frequent occurrence of nanoparticles with diameters close to 20 nm is observed; however, the larger nanoparticles are expected to dominate the absorption and scattering of light. The larger particles may form due to amalgamation of smaller particles whereas the smaller particles are formed by diffusion; this could explain the bimodal size distribution.

Diffraction mode TEM was also applied. In figure 4, the diffraction pattern of sample 1 is depicted. The rings in the pattern consist of many small bright dots because the nanoparticles are crystalline and randomly orientated. The radii of the rings were matched with crystallography data for the metallic β-Sn structure using the calibrated camera constant of the TEM; the crystallographic directions are also denoted on figure 4. Furthermore, the presence of neither α-Sn nor tin-oxides was indicated. Diffraction mode TEM on all samples revealed that the nanoparticles are crystalline with a metallic β-Sn crystal structure. After annealing, however, the nanoparticles have a higher degree of crystallinity.

The tin nanoparticles of the a-Si host sample were also examined using TEM; figure 5 shows a plane-view bright field image of the film that was heat treated at 400°C. The dark contrast reveal spherical tin nanoparticles in the film of an average diameter of 10±3 nm. Nanoparticles were also found before heat treatment and after heat treatment at 700°C; the average diameters were then 8±4 and 12±7 nm, respectively. Diffraction mode TEM showed that the silicon turned polycrystalline only after the heat treatment at 700°C.

*3.2 Results from optical spectroscopy*

Figure 6 depicts extinction spectra of sample one to four, and of a quartz substrate. As indicated by the arrows the extinction of sample two, three and four peak at 5.55, 5.50 and 5.35 eV, respectively. Thus, as the particle size increase, the peak position red shifts. Sample one, however, does not have a pronounced peak, and the peak of sample two is very broad; increasing the particle size further, however, narrows the peak width. A possible concern could be oxidation of Sn over time via micro paths in the sputtered $SiO_2$. Therefore, the extinction measurements were repeated after a few months; however, the extinction properties were permanent indicating no oxidation with time. The extinction of the quartz substrate does not change significantly with energy and is always below 0.06, corresponding to more than 87% transmission.

The effects of absorption in the $SiO_2$ host film were neglected in the analysis above because $SiO_2$ has very low absorption in the considered energy range. When the particles are embedded in a-Si, however, host absorption was found substantial and must be taken into consideration. The optical properties of a-Si depend strongly on the deposition conditions [15], therefore a more detailed investigation and analysis is required. In figure 7a, the absorption spectra of the tin nanoparticles in a-Si and the corresponding reference samples are depicted. These spectra were derived from reflection and transmission spectra measured as specified in the experimental section and using equation (1).

The optical absorption edge in a-Si is obvious on figure 7a; the absorption of all the samples containing a-Si layers increase between 1.5 and 2.0 eV by an amount of about 0.4 (a.u.). The absorption of the samples containing Sn is higher than the absorption of the references; this must be a consequence of the presence of tin in the sample. Note, however, that both tin distributed in the film and in the form of nanoparticles can contribute to the higher absorption. In order to examine this further, the absorption difference was calculated by subtracting the a-Si reference absorption from the sample absorption; the result is shown in figure 7b. This approach assumes that the absorption in the a-Si is independent of whether tin in present in the sample or not. A theoretical account of this analysis method is given in Ref. [19]. The difference in film thickness and roughness between sample and reference may affect the absorption difference spectra, but the effect is not likely to be distinctly energy dependent and thus will not contribute to the appearance of peaks in the spectra.

On graph figure 7b, two peaks are pointed out by arrows; peak P1 is centered at 1.8 eV and peak P2 is centered at 2.5 eV. For the samples that were not heat treated the peak P2 is not present; however, it is observed for heat treatment at 400°C and even more distinctly at 700°C. The irregularity in the spectra of figure 7a and 7b near 1.25 and 1.4 eV is a consequence of changing the grating and the detector in the spectrometer at 1000 and 900 nm.

## 4 Discussion

The host material properties have direct impact on the LSP resonance of the nanoparticles. In the Rayleigh limit, i.e. particles small compared to the wavelength of the light, the LSP resonance energy $\omega_{lsp}$ for spherical nanoparticles embedded in an infinite, uniform host material can be approximated by [20]:

$$\omega_{lsp} = \frac{\omega_p}{\sqrt{\varepsilon_{inter} + 2\varepsilon_{host}}} \qquad (2)$$

where $\varepsilon_{host}$ is the dielectric constant of the surrounding medium, $\varepsilon_{inter}$ is the constant interband screening in the metal, and $\omega_p$ is the bulk plasmon energy of the metal. The bulk plasmon energy of metallic tin was predicted to be 8.25 eV and 9.20 eV for light polarized perpendicular and parallel to the c-axis of tins tetragonal unit cell, respectively [21]. Using this with $\varepsilon_{SiO_2} = 1.5^2$ and $\varepsilon_{inter} = 1$ in equation (2) results in a LSP resonance energy of 3.5 and 3.9 eV, respectively; however, a more detailed calculation including the interband dielectric function of β-Sn was conducted in reference [21] resulted in $\omega_{lsp}$ = 6.0 and 6.25 eV for small spherical particles in a SiO$_2$ host, again, depending on the crystallographic direction in metallic tin. The latter method is anticipated to be most accurate since it takes into account the energy dependence of the screening properties of tin. The detailed estimate of the LSP resonance energies for small Sn particles in the SiO$_2$ host are in acceptable agreement with the observed energies around 5.5 eV.

In addition to the surrounding medium, the size of the nanoparticles also has an effect on the LSP resonance energy. The tendency observed in the literature for gold and silver nanoparticles is that increasing the particle size cause a red shift in the LSP resonance [5, 7, 11]. However, also a blue shift was observed for particles with a diameter of a few nanometers and less [11]. Here the particles were larger than a few nanometers, and a red shift was observed which is in agreement with observations in references [5, 7, 11]. Depending on the model used, different theories predict either a red shift or a blue shift with increasing particle size [10]. In fact, a full calculation of LSP resonances using Mie theory [1] predicts a much larger red shift with size than the one observed here. A possible explanation is that our particles are not surrounded by an infinite SiO$_2$ host but rather by a relatively thin shell as shown in figure 2. Hence, the screening of the plasmons by the surroundings is reduced by the presence of voids. In the Rayleigh limit of small particles, the plasmon resonance of metallic particles embedded in a dielectric shell with dielectric constant $\varepsilon_{host}$ and otherwise in vacuum is easily calculated using the method of reference [22]. Thus, if the ratio between outer and inner radius of the shell is $f$ a resonance is predicted at the energy

$$\omega_{lsp} = \omega_p \sqrt{\frac{2(\varepsilon_{host}-1) + f^3(\varepsilon_{host}+2)}{2(\varepsilon_{host}-1)(\varepsilon_{inter}-\varepsilon_{host}) + f^3(\varepsilon_{host}+2)(\varepsilon_{inter}+2\varepsilon_{host})}}. \qquad (3)$$

This expression interpolates between screening by the homogeneous host material for $f = \infty$ and vanishing screening by the surroundings for $f = 1$. It follows that the resonance lies somewhere between the fully screened and unscreened values. Increasing the particle size will cause a blue shift in $\omega_{lsp}$, this is easily shown using eq. 3 with a constant thickness $t$ of the surrounding $SiO_2$ layer and thus $f = (t+r)/r$, where $r$ is the particle radius. This explains the observed reduction in the red shift. Besides the effect of reduced screening due to the voids, the ellipsoidal shape of the particles observed in sample 4 can cause a red shift in the LSP peak position [20]. This may explain the extra red shift of sample 4 compared with samples 2 and 3. Furthermore, the close proximity of nanoparticles in our samples may also influence plasmon resonances. Generally, particles should be separated by a distance of at least their diameter to behave as individual scatterers. Hence, for our samples treating particles as independent is only approximately correct. Regardless of the theoretical method, the LSP resonance of small and medium sized metallic tin nanoparticles in $SiO_2$ is expected to be in the ultraviolet where it was also observed. Altogether, the size, shape and screening effects described above can explain the variation in LSP peak position observed, and we believe the particles size is the most decisive of these parameters.

Equations (2) and (3) are based on the excitation of dipole LSPs; but also higher order modes can be excited, especially in larger particles. The consequence of higher order excitations is broader resonance lines. The broad size distribution of the nanoparticles can also cause broader peaks since the LSP resonance is size dependent. The directional anisotropy of β-Sn is an additional reason to expect broad peaks; however, this effect is independent of nanoparticle size. Furthermore, the standard deviation in the nanoparticle size distribution is increasing with increasing particle size which might also cause the peak to broaden. Surprisingly, the width of the peak is narrowing with increased particle size. A possible explanation is that the scattering of electrons on the particle surface has less influence for larger particles.

For the system consisting of β-Sn nanoparticles surrounded by homogeneous a-Si, two peaks (P1 and P2) were observed in the absorption difference spectra. These two peaks are related to the presence of tin since this is the only difference between sample and reference. The positions of the peak P1, and of the optical absorption edge of a-Si are both centered at 1.8 eV. We believe P1 stems from tin altering the optical absorption edge presumably by increasing the disorder in a-Si.

Using equation (2) with $\varepsilon_{a-Si} = 3.4^2$ and $\varepsilon_{inter} = 1$ gives a LSP resonance energy of 1.7 eV and 1.9 eV. However, a full calculation including interband screening [21] predicts the LSP resonance to be at 2.4-2.6 eV when the diameter is 10 nm. This is in very good agreement with the observed peak P2 centered at 2.5 eV, therefore, the peak P2 is assigned to the LSP resonance in tin surrounded by a-Si. Furthermore, it was observed that the intensity of the P2 peak increases with increasing heat treatment temperature. Two possible explanations for this correlation are the formation of larger nanoparticles due to Oswald ripening and the increased crystallinity of the host matrix during heat treatment.

Nanoparticles of gold and silver were previously used to enhance photovoltaic devices by utilizing LSP resonances in [7-9]. In thin film silicon solar cells, the most interesting wavelength region to enhance is from 800 to 1100 nm where the photo-current generation is inefficient because of the indirect band-gap of silicon. Consequently, among the materials considered here a-Si is the best candidate for a tin nanoparticle host material. On the other hand, $SiO_2$ may be useful as host in UV related applications.

## 5 Conclusion

The behavior of the LSP resonance of β-Sn nanoparticles in $SiO_2$ and a-Si was investigated. After heat treatment, sputtering deposition of trilayered $SiO_2$/Sn/$SiO_2$ thin films resulted in spherical, crystalline, β-Sn nanoparticles due to thermal precipitation. The nanoparticle diameter was controlled by varying the content of Sn in the films. Thus β-Sn nanoparticles embedded in $SiO_2$ with average diameters ranging from 15 to 80 nm were produced. The LSP resonance of these particles was observed by optical spectroscopy around 5.5 eV, and it was found that the resonance shifts towards lower energy when the nanocrystal size increases. Also β-Sn nanoparticles embedded in a-Si were produced and characterized. By analyzing the change in absorption caused by the presence of the nanoparticles, the LSP resonance was found at 2.5 eV.

**Acknowledgements**

The authors would like to thank Pia Bomholt for preparation of the TEM samples. Duncan Sutherland is acknowledged for providing the spectrometer and for fruitful discussions.

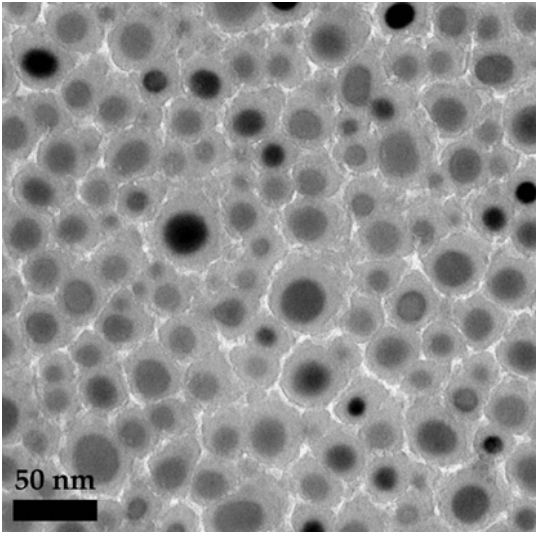

**Figure 1**. TEM plane-view micrograph of sample 1 before the heat treatment.

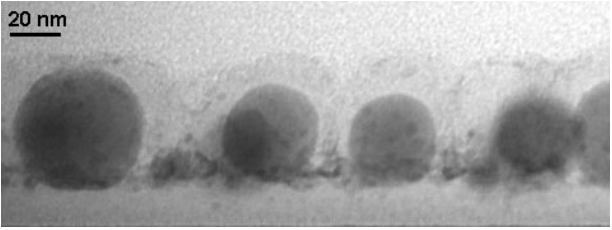

**Figure 2.** TEM cross-section micrograph showing nanoparticles in sample 2.

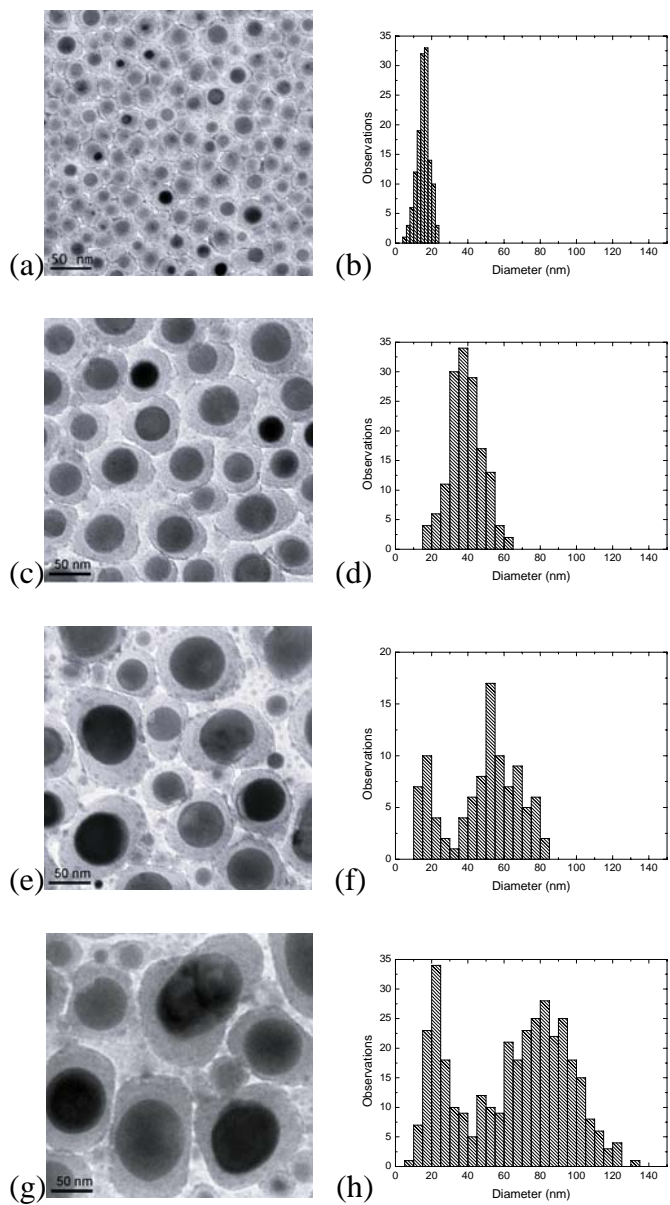

**Figure 3.** TEM plane-view micrographs with corresponding nanoparticle size distribution for sample 1 (a, b), sample 2 (c, d), sample 3 (e, f) and sample 4 (g, h).

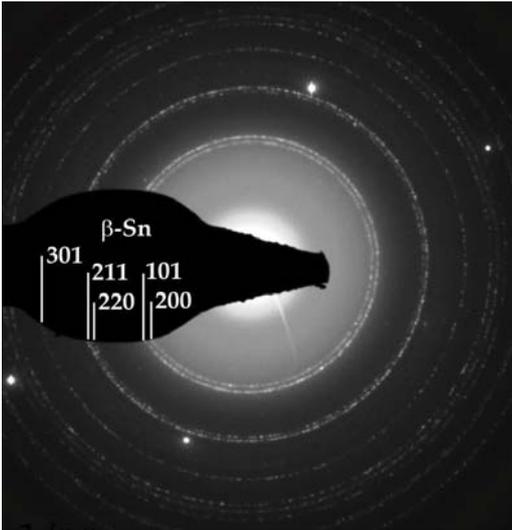
**Figure 4.** Diffraction mode TEM image of sample 1 after the heat treatment.

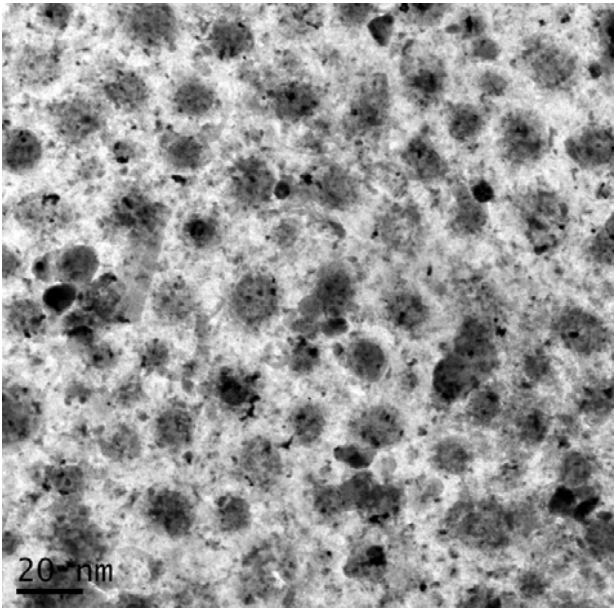

**Figure 5.** TEM plane-view micrograph of a Si/Sn/Si sample that was heat treated at 400°C.

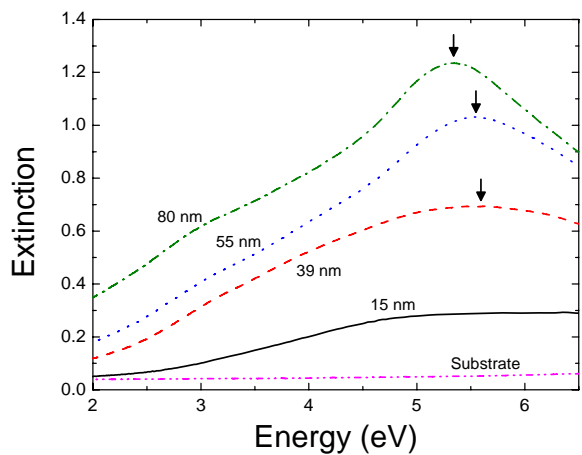

**Figure 6.** Extinction spectra of the substrate, and of metallic tin nanoparticles in SiO$_2$ with diameter: 15, 39, 55, and 80 nm. The extinction maxima at 5.55, 5.50, and 5.35 eV are indicated by an arrow.

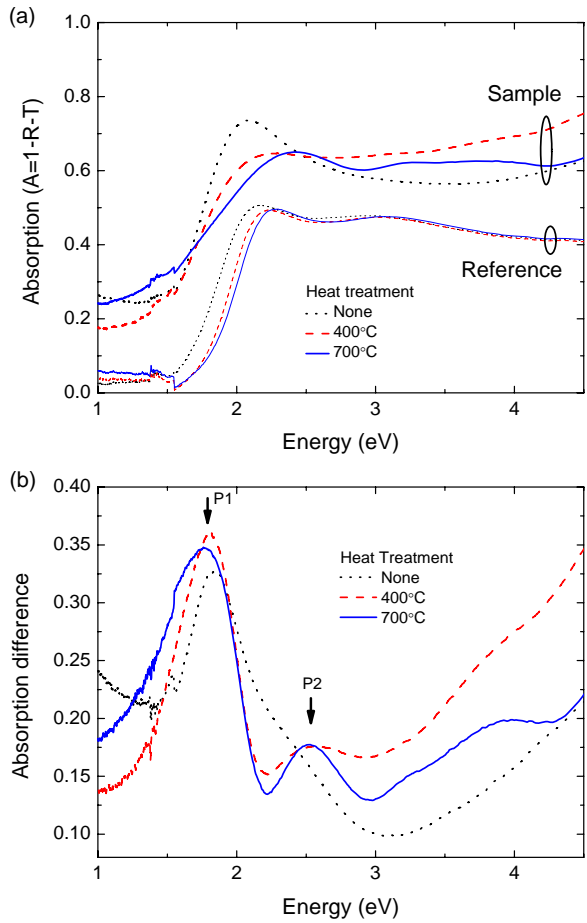

**Figure 7.** (a) Absorption spectra of metallic tin nanoparticles in a-Si samples and in reference samples without tin. Dotted curves, not heat treated; dashed curves heat treated at 400°C; and solid curves heat treated at 700°C. (b) Spectra of difference in absorption with two peaks indicated by two arrows. Peak P1 is centered at 1.8 eV, and peak P2 is centered at 2.5 eV.

**Table 1.** Summary of sample properties: Sn content, nanoparticle diameter with standard deviation, and extinction-maxima.

|  | Tin areal density ($10^{16}$ atoms/cm$^2$) | Thickness in β-tin equivalent (nm) | Average nanoparticle diameter (nm) | Extinction maximum (eV) |
|---|---|---|---|---|
| Sample 1 | 1.7 | 4.5 | 15 ± 4 | - |
| Sample 2 | 3.5 | 9.5 | 39 ± 9 | 5.55 |
| Sample 3 | 5.0 | 13.5 | 16 ± 5 and 55 ± 12 | 5.50 |
| Sample 4 | 5.9 | 16.0 | 22 ± 8 and 80 ± 30 | 5.35 |